\documentclass{PoS}

\def\mathswitch#1{\relax\ifmmode#1\else$#1$\fi}
\def\mathswitchr#1{\relax\ifmmode{\mathrm{#1}}\else$\mathrm{#1}$\fi}

\newcommand{\als}{\alpha_{\mathrm{s}}}

\newcommand{\GF}{\mathswitch {G_\mu}}
\newcommand{\sw}{\mathswitch {s_\rw}}
\newcommand{\rT}{{\mathrm{T}}}
\newcommand{\rj}{\mathrm{j}}
\newcommand{\rw}{\mathrm{w}}
\newcommand{\gammainduced}{\mbox{\scriptsize $\gamma$-induced}}
\newcommand{\si}{\sigma}
\newcommand{\de}{\delta}
\newcommand{\ppjjh}{\Pp\Pp\to\PH+2\mathrm{jets}+X}

\newcommand{\PH}{\mathswitchr H}
\newcommand{\PW}{\mathswitchr W}
\newcommand{\PZ}{\mathswitchr Z}
\newcommand{\Pp}{\mathswitchr p}

\def\mathswitch#1{\relax\ifmmode#1\else$#1$\fi}
\newcommand{\MH}{\mathswitch {M_\PH}}
\newcommand{\MW}{\mathswitch {M_\PW}}
\newcommand{\MZ}{\mathswitch {M_\PZ}}

\newcommand{\GeV}{\unskip\,\mathrm{GeV}}
\newcommand{\fb}{\unskip\,\mathrm{fb}}

\def\reffi#1{\mbox{Figure~\ref{#1}}}

\def\refta#1{\mbox{Table~\ref{#1}}}
\def\citere#1{\mbox{Ref.~\cite{#1}}}
\def\citeres#1{\mbox{Refs.~\cite{#1}}}

\title{Electroweak and QCD corrections to
Higgs-boson production in vector-boson fusion at the LHC}

\ShortTitle{Electroweak and QCD corrections to
Higgs production in VBF at the LHC}

\author{Mariano Ciccolini,$^a$ \speaker{Ansgar Denner}$^a$ and Stefan
  Dittmaier$^{b,c}$
\thanks{ This work is supported in part by the European
Community's Marie-Curie Research Training Network HEPTOOLS under
contract MRTN-CT-2006-035505.}\\
\llap{$^a$}Paul Scherrer Institut, W\"urenlingen und Villigen,
CH-5232 Villigen PSI, Switzerland \\
\llap{$^b$}Max-Planck-Institut f\"ur Physik
(Werner-Heisenberg-Institut), D-80805 M\"unchen, Germany\\
\llap{$^c$}Faculty of Physics, University of Vienna,
A-1090 Vienna, Austria\\
E-mail: \email{Mariano.Ciccolini@psi.ch}, \email{Ansgar.Denner@psi.ch}, \email{Stefan.Dittmaier@mppmu.mpg.de}}

\abstract{  Radiative corrections of strong and electroweak interactions are
  presented at next-to-leading order for Higgs-boson production in the
  weak-boson-fusion channel 
  at the LHC.  The calculation includes all weak-boson fusion and
  quark--antiquark annihilation diagrams as well as all related
  interferences.  The electroweak corrections, which also include real
  corrections from incoming photons and leading heavy-Higgs-boson
  effects at two-loop order, are of the same size as the QCD
  corrections, viz.\ typically at the level of $5{-}10\%$ for a
  Higgs-boson mass up to $\sim700\GeV$.  In general, they do not
  simply rescale differential distributions, but induce distortions at
  the level of 10\%.  The discussed corrections have been implemented
  in a flexible Monte Carlo event generator.}

\FullConference{8th International Symposium on Radiative Corrections (RADCOR)\\
                 October 1-5 2007\\
                 Florence, Italy}

\begin{document}

\section{Introduction}

The electroweak (EW) production of a Standard Model Higgs boson in
association with two hard jets in the forward and backward regions of
the detector---frequently quoted as ``vector-boson fusion'' (VBF)---is
a cornerstone in the Higgs search both in the ATLAS \cite{Asai:2004ws}
and CMS \cite{Abdullin:2005yn} experiments at the LHC and also plays
an important role in the determination of Higgs couplings at this
\looseness -1
collider.

Higgs+2jets production in pp collisions proceeds through two
different channels.
The first channel corresponds to a pure EW process. It comprises the
scattering of two (anti-)quarks mediated by $t$- and $u$-channel W- or
Z-boson exchange, with the Higgs boson radiated off the virtual weak
boson.  It also involves Higgs-boson radiation off a W- or Z-boson
produced in $s$-channel quark--antiquark annihilation (Higgs-strahlung
process), with the weak boson decaying hadronically.
The second channel proceeds through strong interactions, the
Higgs boson being radiated off a heavy-quark loop that couples to any
parton of the incoming hadrons via gluons
\cite{DelDuca:2001fn}.

In the weak-boson-mediated processes, the two scattered quarks are
usually visible as two hard forward jets, in contrast to other jet
production mechanisms, offering a good background suppression
(transverse-momentum and rapidity cuts on jets, jet rapidity gap,
central-jet veto, etc.).  Applying appropriate event selection
criteria (see e.g. \citere{Barger:1994zq} and references in
\citeres{Spira:1997dg,Djouadi:2005gi}) it is possible to sufficiently
suppress background and to enhance the VBF channel over the hadronic
Higgs+2jets production mechanism.

In order to match the required precision for theoretical predictions
at the LHC, QCD and EW corrections are needed. When VBF cuts are
imposed, the cross section can be approximated by the contribution of
squared $t$- and $u$-channel diagrams only, which reduces the QCD
corrections to vertex corrections to the weak-boson--quark coupling.
Explicit next-to-leading order (NLO) QCD calculations in this
approximation exist since more than a decade
\cite{Spira:1997dg,Han:1992hr}, while corrections to distributions
have been calculated in the last few years
\cite{Figy:2003nv,Figy:2004pt}.  Recently, the full NLO EW and QCD
corrections to this process have become available
\cite{Ciccolini:2007jr,Ciccolini:2007ec}. This calculation includes,
for the first time, the complete set of EW and QCD diagrams, namely
the $t$-, $u$-, and $s$-channel contributions, as well as all
interferences.

In these proceedings we briefly summarize the details of the NLO
EW and QCD calculation and give some new results on distributions for
a Higgs-boson mass of $200\GeV$. Numerical results for the Higgs-mass
dependence of the total cross section with and without VBF cuts as
well as distributions for $\MH=120\GeV$ have been presented in
\citeres{Ciccolini:2007jr,Ciccolini:2007ec}.

\section{Details of the NLO calculation}

We have calculated the complete QCD and EW NLO corrections to Higgs
\looseness -1
production via weak VBF at the LHC.  At LO, this process receives
contributions from the partonic processes $qq\to\PH qq$,
$q\bar{q}\to\PH q\bar{q}$, and $\bar{q}\bar{q}\to\PH\bar{q}\bar{q}$.
For each relevant configuration of external quark flavours one or two
Feynman diagrams contribute in LO.  All LO and one-loop NLO diagrams
are related by crossing symmetry to the corresponding decay amplitude
$\PH\to q\bar{q}q\bar{q}$. The QCD and EW NLO corrections to these
decays were discussed in detail in \citere{Bredenstein:2006rh}, where
a representative set of Feynman diagrams can be found.

At NLO, there are about 200 EW one-loop diagrams
per tree diagram in each flavour channel.
The calculation of the one-loop diagrams has been performed in the
conventional 't~Hooft--Feynman gauge and in the background-field
formalism using the conventions of \citeres{Denner:1991kt} and
\cite{Denner:1994xt}, respectively. The masses of the external
fermions have been neglected whenever possible, i.e.\ everywhere but
in the mass-singular logarithms.
The amplitudes have been generated with {\sl FeynArts}, using the two
independent versions 1 \cite{Kublbeck:1990xc} and 3
\cite{Hahn:2000kx}.  The algebraic evaluation has been performed in
two completely independent ways. One calculation is based on the
in-house {\sl Mathematica} program that was already used in the
algebraic reduction of NLO corrections to the $\PH\to4\,$fermions
decays \cite{Bredenstein:2006rh}.  The other has
been completed with the help of {\sl FormCalc} \cite{Hahn:1998yk}.

In the $s$-channel diagrams intermediate W and Z~bosons can become
resonant, corresponding to $\PW\PH/\PZ\PH$ production with subsequent
gauge-boson decay. In order to consistently include these resonances,
we use the 
``complex-mass scheme'', which was introduced in
Ref.~\cite{Denner:1999gp} for LO calculations and generalized to the
one-loop level in Ref.~\cite{Denner:2005fg}. In this approach the W-
and Z-boson masses are consistently considered as complex quantities,
defined as the locations of the propagator poles in the complex plane.
The scheme respects all relations that follow from gauge
invariance.

The tensor integrals are evaluated as in the calculation of the
NLO corrections to ${\rm e}^+{\rm e}^-\to4\,$fermions
\cite{Denner:2005fg,Denner:2005es}. They are recursively reduced to
master integrals at the numerical level. The scalar master integrals
are evaluated for complex masses using the methods and results of
\citere{'tHooft:1978xw}. 
Tensor and scalar 5-point functions are directly expressed in terms of
4-point integrals \cite{Denner:2002ii}.  Tensor 4-point and 3-point
integrals are reduced to scalar integrals with the Passarino--Veltman
algorithm \cite{Passarino:1978jh} as long as no small Gram determinant
appears in the reduction. If small Gram determinants occur, the
alternative schemes described in Ref.~\cite{Denner:2005nn} are
applied. 

Real QCD corrections consist of gluon emission and processes with $gq$
and $g\bar q$ initial states. Analogously, real photonic corrections
comprise photon bremsstrahlung and photon-induced processes with
$\gamma q$ and $\gamma\bar q$ initial states.  The matrix elements for
these corrections have been evaluated using the Weyl--van der Waerden
spinor technique as formulated in Ref.~\cite{Dittmaier:1998nn} and
have been checked against results obtained with {\sl Madgraph}
\cite{Stelzer:1994ta}.  The phase-space integration is performed using
multi-channel Monte Carlo techniques \cite{Hilgart:1992xu} implemented
in different ways in two different generators.

All types of real corrections involve singularities from collinear
initial-state splittings which are regularized with small quark
masses.  The mass singularities are absorbed via factorization by the
usual PDF redefinition both for the QCD and photonic corrections (see,
e.g., \citere{Diener:2005me}).  Technically, the soft and collinear
singularities for real gluon or photon emission are isolated both in
the dipole subtraction method following \citere{Dittmaier:1999mb} and
in the phase-space slicing method. For gluons or photons in the
initial state the subtraction and slicing variants described in
\citere{Diener:2005me} are applied. The results presented in the
following are obtained with the subtraction method, which numerically
performs better.

\section{Numerical results}

We use the input parameters as given in \citere{Ciccolini:2007ec}.
Since quark-mixing effects are suppressed, the CKM matrix is set to
the unit matrix.  The electromagnetic coupling is fixed in the $G_\mu$
scheme, i.e.\ it is set to $\alpha_{\GF}=\sqrt{2}\GF\/\MW^2\sw^2/\pi$,
because this accounts for electromagnetic running effects and some
universal corrections of the $\rho$ parameter. 
We use the MRST2004QED PDF \cite{Martin:2004dh} which consistently
include ${\cal O}(\alpha)$ QED corrections and a photon distribution
function for the proton. We use only four quark flavours for the
external partons, i.e.\ we do not take into account the contribution
of bottom quarks, which is suppressed.  The renormalization and
factorization scales are set to $\MW$, 5 flavours are included in the
two-loop running of $\als$, and $\als(\MZ)=0.1187$.  We apply typical
VBF cuts to the outgoing jets as described in detail in
\citere{Ciccolini:2007ec}.
 

In \refta{ta:xsection_cuts} we present integrated cross sections for
$\MH=120$, 150, 200, 400, and $700\GeV$ for VBF cuts.  We list the LO
cross section $\sigma_{\mathrm{LO}}$, the cross section
$\sigma_{\mathrm{NLO}}$ including QCD+EW corrections, and the relative
QCD and EW corrections, $\delta_{\mathrm{QCD}}$ and
$\delta_{\mathrm{EW}}$, respectively.  The complete EW corrections
$\delta_{\mathrm{EW}}$ also comprise the corrections from
photon-induced processes $\delta_{\gammainduced}$, which turn out to
be $\sim+1\%$.  The QCD corrections are dominated by the known
(vertex-like) corrections to the squared $t$- and $u$-channel VBF
diagrams, while corrections to interference terms are
below $0.1\%$ (see also \citere{Andersen:2006ag}).%
\begin{table}
 \def\phm{\phantom{-}}
\def\phn{\phantom{0}}
\def\phe{\phantom{(0)}}
\centerline{
\begin{tabular}{|c|c|c|c|c|c|}
\hline
$\MH\ [\GeV]$ & 120 & 150 & 200 & 400& 700\\
\hline
$\si_{\mathrm{LO}}\ [\fb]$ 
& 1876.3(5) %
& 1589.8(4) %
& 1221.1(3) %
&  487.31(9)
& 160.67(2) %
\\
$\si_{\mathrm{NLO}}\ [\fb]$ 
& 1665(1) %
& 1407.5(8) %
& 1091.3(5) %
&  435.4(2)
&  160.36(5) %
\\
$\de_{\mathrm{EW}}\ [\%]$ 
& $-6.47(2)$  %
& $-6.27(2)$  %
& $-4.98(1)$  %
& $-3.99(1)$  %
& $\phm 6.99(2)$  %
\\
$\de_{\gammainduced}\ [\%]$ 
& $\phm1.10\phe$ 
& $\phm1.15\phe$ 
& $\phm1.22\phe$ 
& $\phm1.38\phe$ 
& $\phm1.55\phe$ 
\\
$\de_{\mathrm{QCD}}\ [\%]$ 
& $-4.77(4)$  %
& $-5.20(4)$  %
& $-5.65(3)$ %
& $-6.67(3)$ 
& $-7.18(2)$
\\
\hline
\end{tabular}
}
\caption{Cross sections for $\ppjjh$ in LO and NLO with VBF cuts
  and relative EW and  QCD corrections.
The contribution
  $\delta_{\gammainduced}$ from $\gamma$-induced processes (which
  is part of $\delta_{\mathrm{EW}}$) is also given separately.}
\label{ta:xsection_cuts}
\end{table}

\begin{figure}
  \includegraphics[bb= 85 440 285 660, scale=1]{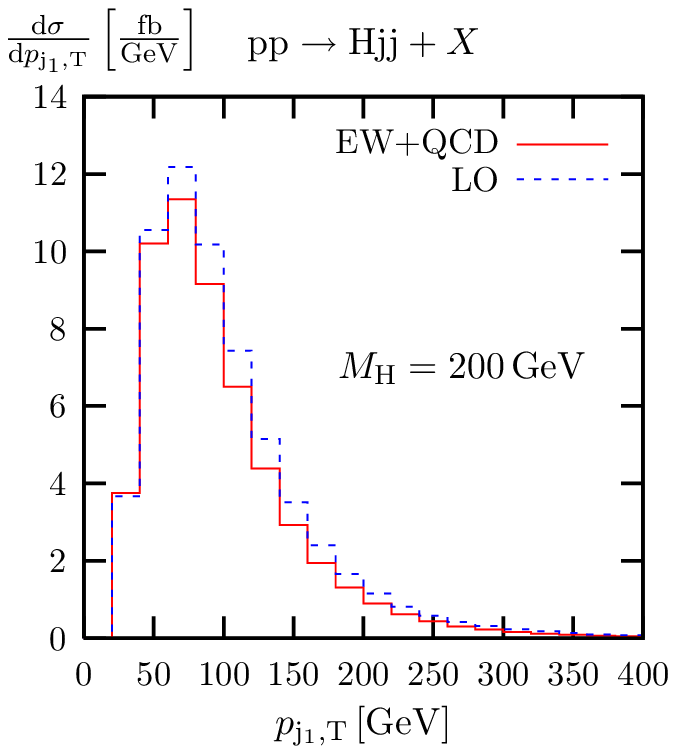} \quad
  \includegraphics[bb= 85 440 285 660, scale=1]{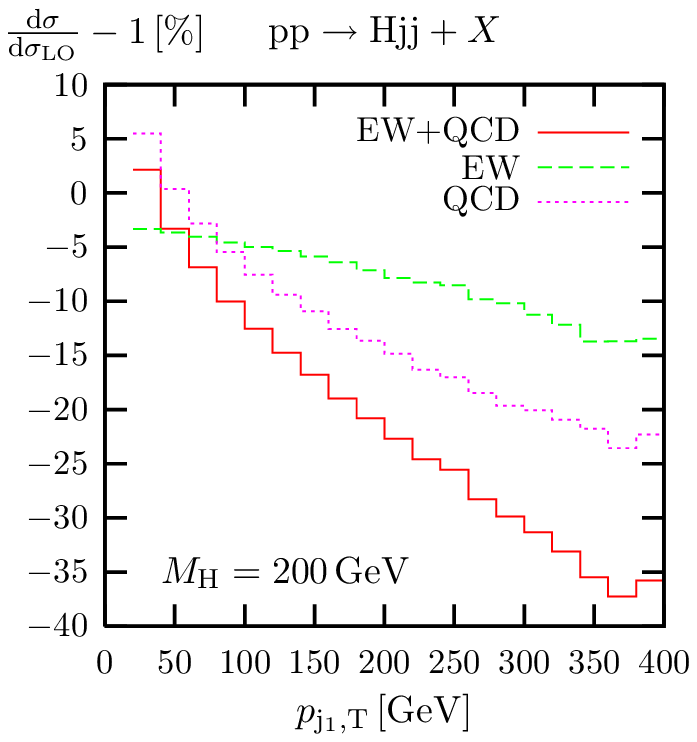}
\caption{Distribution in the transverse momentum
  $p_{\mathrm{j_1},\rT}$ of the harder tagging jet (left) and
  corresponding relative corrections (right) for $\MH=200\GeV$.}
\label{fi:pt1}
\vspace{2em}
  \includegraphics[bb= 85 440 285 660, scale=1]{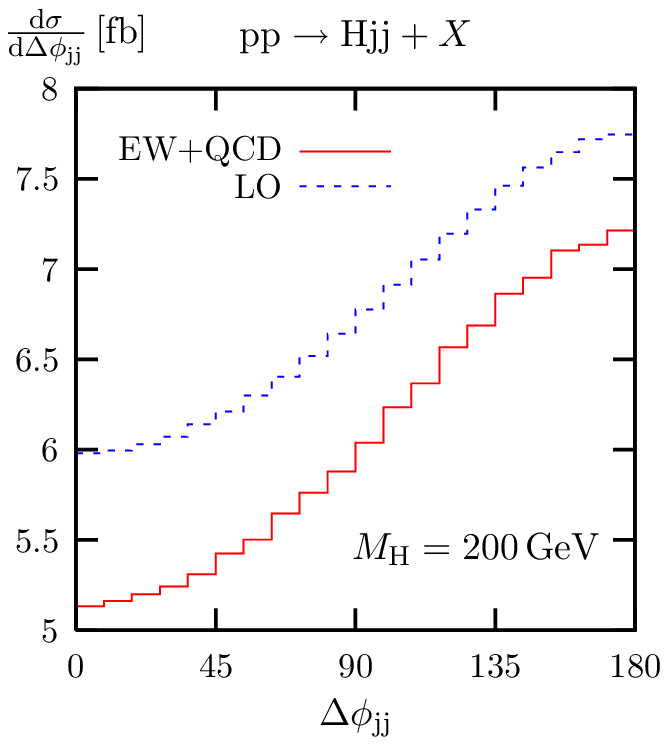}
  \quad \includegraphics[bb= 85 440 285 660,
  scale=1]{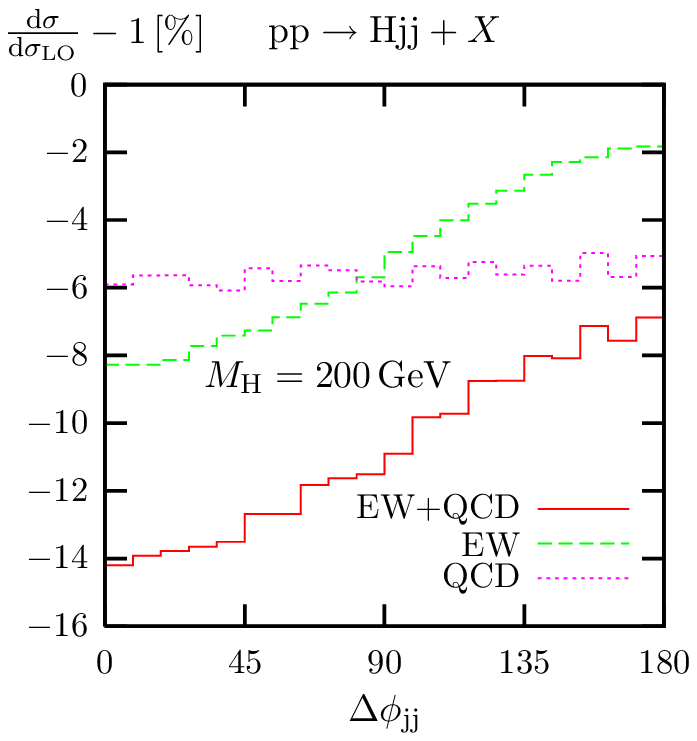}
\caption{Distribution in the azimuthal angle difference $\Delta
  \phi_{\mathrm{jj}}$ of the tagging jets (left) and corresponding
  relative corrections (right) for $\MH=200\GeV$.}
\label{fi:phi}
\end{figure}

The EW corrections to distributions for $\MH=200\GeV$ are
qualitatively similar to the distributions for $\MH=120\GeV$ presented
in \citere{Ciccolini:2007ec}.  As an example we show the distribution
in the transverse momentum $p_{\mathrm{j}_1,\rT}$ of the harder
tagging jet $\mathrm{j}_1$ (jet with highest $p_{\rT}$ passing all
cuts) in \reffi{fi:pt1}.  We plot the absolute predictions in LO and
in NLO including QCD and EW corrections. In addition, we show the
relative QCD and EW corrections separately, as well as their sum.
QCD and EW corrections become more and more negative with increasing
$p_\mathrm{j_1,T}$. For low transverse momentum these corrections are at
the level of 5\%, while for $p_{\rj_1,\rT}=400\GeV$ they add up to about
$-40\%$. This induces a substantial change in shape of this
distribution.

In \reffi{fi:phi} we present the distribution in the azimuthal angle
separation of the two tagging jets. This distribution is particularly
sensitive to non-standard contributions to the $\PH VV$ vertices
\cite{Figy:2004pt}.  As expected for VBF processes, there is a large
azimuthal angle separation between the two tagging jets.  While QCD
corrections are almost flat in this variable, the QCD+EW corrections
exhibit a dependence on $\Delta \phi_{\mathrm{jj}}$ on the level of
7\%, which is almost twice as big as for $\MH=120\GeV$
\looseness -1
\cite{Ciccolini:2007ec}.

\section{Conclusions}

Radiative corrections of strong and electroweak interactions have been
discussed at next-to-leading order for Higgs production via
vector-boson fusion at the LHC.  The electroweak corrections affect
the cross section by $5\%$, and are thus as important as the QCD
corrections in this channel. They do not simply rescale distributions
but induce distortions at the level of 10\%.


\begin{thebibliography}{99}
\newcommand{\ap}[3]{{\sl Ann.~Phys.} {\bf #1} (19#2) #3}
\newcommand{\zp}[3]{{\sl Z.~Phys.} {\bf #1} (19#2) #3}
\newcommand{\np}[3]{{\sl Nucl.~Phys.} {\bf #1} (19#2) #3}
\newcommand{\pl}[3]{{\sl Phys.~Lett.} {\bf #1} (19#2) #3}
\newcommand{\pr}[3]{{\sl Phys.~Rev.} {\bf #1} (19#2) #3}
\newcommand{\prl}[3]{{\sl Phys.~Rev.~Lett.} {\bf #1} (19#2) #3}
\newcommand{\fp}[3]{{\sl Fortschr.~Phys.} {\bf #1} (19#2) #3}
\newcommand{\jp}[3]{{\sl J.~Phys.} {\bf #1} (19#2) #3}
\newcommand{\cpc}[3]{{\sl Comput.~Phys.~Commun.} {\bf #1} (19#2) #3}
\newcommand{\ijmp}[3]{{\sl Int.~J.~Mod.~Phys.} {\bf #1} (19#2) #3}
\newcommand{\nim}[3]{{\sl Nucl.~Instr.~Meth.} {\bf #1} (19#2) #3}
\newcommand{\nc}[3]{{\sl Nuovo Cimento} {\bf #1} (19#2) #3}
\newcommand{\mpl}[3]{{\sl Modern Physics Letters} {\bf #1} (19#2) #3}
\newcommand{\vj}[4]{{\sl #1} {\bf #2} (19#3) #4}

\bibitem{Asai:2004ws}
  S.~Asai {\it et al.},
  \emph{Eur.\ Phys.\ J.\  C} {\bf 32S2} (2004) 19
  [hep-ph/0402254].

\bibitem{Abdullin:2005yn}
  S.~Abdullin {\it et al.},
  \emph{Eur.\ Phys.\ J.\  C} {\bf 39S2} (2005) 41.




\bibitem{DelDuca:2001fn}
  V.~Del Duca {\it et al.},
  \emph{Nucl.\ Phys.\  B} {\bf 616} (2001) 367
  [hep-ph/0108030];\\
%
  J.~M.~Campbell, R.~K.~Ellis and G.~Zanderighi,
  \emph{JHEP} {\bf 0610} (2006) 028
  [hep-ph/0608194].

\bibitem{Barger:1994zq}
  V.~D.~Barger, R.~J.~N.~Phillips and D.~Zeppenfeld,
  \emph{Phys.\ Lett.\  B} {\bf 346} (1995) 106
  [hep-ph/9412276];\\
%
  D.~L.~Rainwater and D.~Zeppenfeld,
  \emph{JHEP} {\bf 9712} (1997) 005
  [hep-ph/9712271];\\
%
  D.~L.~Rainwater, D.~Zeppenfeld and K.~Hagiwara,
  \emph{Phys.\ Rev.\  D} {\bf 59} (1999) 014037
  [hep-ph/9808468];\\
%
  D.~L.~Rainwater and D.~Zeppenfeld,
  \emph{Phys.\ Rev.\  D} {\bf 60} (1999) 113004
  [Erratum-ibid.\  D {\bf 61} (2000) 099901]
  [hep-ph/9906218];\\
%
  V.~Del Duca {\it et al.},
  \emph{JHEP} {\bf 0610} (2006) 016
  [hep-ph/0608158].

\bibitem{Spira:1997dg}
  M.~Spira,
  \emph{Fortsch.\ Phys.}\  {\bf 46} (1998) 203
  [hep-ph/9705337].

\bibitem{Djouadi:2005gi}
  A.~Djouadi,
  hep-ph/0503172.


\bibitem{Han:1992hr}
  T.~Han, G.~Valencia and S.~Willenbrock,
  \emph{Phys.\ Rev.\ Lett.}\  {\bf 69} (1992) 3274
  [hep-ph/9206246].
%
\bibitem{Figy:2003nv}
  T.~Figy, C.~Oleari and D.~Zeppenfeld,
  \emph{Phys.\ Rev.\  D} {\bf 68} (2003) 073005
  [hep-ph/0306109];\\
%
  E.~L.~Berger and J.~Campbell,
  \emph{Phys.\ Rev.\  D} {\bf 70} (2004) 073011
  [hep-ph/0403194].

\bibitem{Figy:2004pt}
  T.~Figy and D.~Zeppenfeld,
  \emph{Phys.\ Lett.\  B} {\bf 591} (2004) 297
  [hep-ph/0403297];\\

\bibitem{Ciccolini:2007jr}
  M.~Ciccolini, A.~Denner and S.~Dittmaier,
   \emph{Phys.\ Rev.\ Lett.}\ {\bf 99} (2007) 161803
  [arXiv:0707.0381].

\bibitem{Ciccolini:2007ec}
  M.~Ciccolini, A.~Denner and S.~Dittmaier,
  arXiv:0710.4749 [hep-ph], to appear in  Phys.\ Rev.\  D.

\bibitem{Bredenstein:2006rh}
  A.~Bredenstein, A.~Denner, S.~Dittmaier and M.~M.~Weber,
  \emph{Phys.\ Rev.\  D} {\bf 74} (2006) 013004
  [hep-ph/0604011];
%
  \emph{JHEP} {\bf 0702} (2007) 080
  [hep-ph/0611234].


\bibitem{Denner:1991kt}
  A.~Denner,
  \emph{Fortsch.\ Phys.}\  {\bf 41} (1993) 307
  [arXiv:0709.1075 [hep-ph]].

\bibitem{Denner:1994xt}
  A.~Denner, S.~Dittmaier and G.~Weiglein,
  \emph{Nucl.\ Phys.\ B} {\bf 440} (1995) 95
  [hep-ph/9410338].

\bibitem{Kublbeck:1990xc}
J.~K\"ublbeck, M.~B\"ohm and A.~Denner,
\emph{Comput.\ Phys.\ Commun.}\  {\bf 60} (1990) 165;
H.~Eck and J.~K\"ublbeck, {\it Guide to FeynArts 1.0\/},
University of W\"urzburg, 1992.

\bibitem{Hahn:2000kx}
T.~Hahn,
\emph{Comput.\ Phys.\ Commun.}\  {\bf 140} (2001) 418
[hep-ph/0012260].

\bibitem{Hahn:1998yk}
T.~Hahn and M.~P\'erez-Victoria,
\emph{Comput.\ Phys.\ Commun.}\  {\bf 118} (1999) 153
[hep-ph/9807565];\\
%
T.~Hahn,
N\emph{ucl.\ Phys.\ Proc.\ Suppl.}\  {\bf 89} (2000) 231
[hep-ph/0005029].

\bibitem{Denner:1999gp}
A.~Denner, S.~Dittmaier, M.~Roth and D.~Wackeroth,
\emph{Nucl.\ Phys.\ B} {\bf 560} (1999) 33
[hep-ph/9904472].



\bibitem{Denner:2005fg}
  A.~Denner, S.~Dittmaier, M.~Roth and L.~H.~Wieders,
  \emph{Nucl.\ Phys.\  B} {\bf 724} (2005) 247
  [hep-ph/0505042].

\bibitem{Denner:2005es}
  A.~Denner, S.~Dittmaier, M.~Roth and L.~H.~Wieders,
  \emph{Phys.\ Lett.\ B} {\bf 612}, 223 (2005)
  [hep-ph/0502063].

\bibitem{'tHooft:1978xw}
G.~'t Hooft and M.~Veltman,
\emph{Nucl.\ Phys.\ B} {\bf 153} (1979) 365;
%
W.~Beenakker and A.~Denner,
\emph{Nucl.\ Phys.\ B} {\bf 338} (1990) 349;
%
A.~Denner, U.~Nierste and R.~Scharf,
\emph{Nucl.\ Phys.\ B} {\bf 367} (1991) 637.

\bibitem{Denner:2002ii}
  A.~Denner and S.~Dittmaier,
  \emph{Nucl.\ Phys.\ B} {\bf 658} (2003) 175
  [hep-ph/0212259].

\bibitem{Passarino:1978jh}
G.~Passarino and M.~Veltman,
\emph{Nucl.\ Phys.\ B} {\bf 160} (1979) 151.

\bibitem{Denner:2005nn}
  A.~Denner and S.~Dittmaier,
  \emph{Nucl.\ Phys.\ B} {\bf 734} (2006) 62
  [hep-ph/0509141].

\bibitem{Dittmaier:1998nn}
S.~Dittmaier,
\emph{Phys.\ Rev.\ D} {\bf 59} (1999) 016007
[hep-ph/9805445].

\bibitem{Stelzer:1994ta}
  T.~Stelzer and W.F.~Long,
  \emph{Comput.\ Phys.\ Commun.}\  {\bf 81} (1994) 357
  [hep-ph/9401258].

\bibitem{Hilgart:1992xu}
J.~Hilgart, R.~Kleiss and F.~Le Diberder,
\emph{Comput.\ Phys.\ Commun.}\  {\bf 75} (1993) 191.

\bibitem{Diener:2005me}
  K.~P.~Diener, S.~Dittmaier and W.~Hollik,
  \emph{Phys.\ Rev.\  D} {\bf 72} (2005) 093002
  [hep-ph/0509084].

\bibitem{Dittmaier:1999mb}
S.~Dittmaier,
\emph{Nucl.\ Phys.\ B} {\bf 565} (2000) 69
[hep-ph/9904440].

%


\bibitem{Martin:2004dh}
A.~D.~Martin {\it et al.}, 
\emph{Eur.\ Phys.\ J.\ C} {\bf 39} (2005) 155
[hep-ph/0411040].

\bibitem{Andersen:2006ag}
  J.~R.~Andersen and J.~M.~Smillie,
  Phys.\ Rev.\  D {\bf 75} (2007) 037301
  [hep-ph/0611281];\\
%
  J.~R.~Andersen, T.~Binoth, G.~Heinrich and J.~M.~Smillie,
  arXiv:0709.3513 [hep-ph].





  








\end{thebibliography}
\end{document}